\documentclass[aps,prl,twocolumn,superscriptaddress,nofootinbib,showpacs]{revtex4}
\usepackage{slashed}
\usepackage{epsfig,latexsym,cancel,amssymb,amsmath,verbatim,mathrsfs}
\usepackage{color}
\usepackage{graphicx}

\def\ra{\rightarrow}
\def\L{\left(}
\def\R{\right)}
\def\wt{\widetilde}
\def\Ld{\Lambda}
\def\ld{\lambda}
\def\f{\frac}
\newcommand{\be}{\begin{equation}}
\newcommand{\ee}{\end{equation}}
\newcommand{\bea}{\begin{eqnarray}}
\newcommand{\eea}{\end{eqnarray}}
\newcommand{\ba}{\begin{array}}
\newcommand{\ea}{\end{array}}

\long\def\symbolfootnote[#1]#2{\begingroup%
\def\thefootnote{\fnsymbol{footnote}}\footnote[#1]{#2}\endgroup}

\newcommand{\beq}{\begin{equation}}
\newcommand{\eeq}{\end{equation}}
\begin{document}

\title{$H_{u,d}$-messenger Couplings Address the
 $\mu/B_\mu$ \& $A_t/m_{H_u}^2$ Problem and $(g-2)_\mu$ Puzzle}

%\title{Nonradiative EWSB Comes to the Rescue of Gauge Mediated SUSY Breaking: \\
% $\mu/B_\mu$ \& $A_t/m_{H_u}^2-$problem and $(g-2)_\mu$-puzzle}

\author{Zhaofeng Kang}
\email[E-mail: ]{zhaofengkang@gmail.com}
\affiliation{School of Physics, Korea Institute for Advanced Study,
Seoul 130-722, Korea}

%\author{Tao Liu}

%\affiliation{Key Laboratory of Frontiers in Theoretical Physics,
 %            Institute of Theoretical Physics,  Chinese Academy of Sciences,
  %           Beijing 100190,  P. R. China }

\date{\today}

\begin{abstract}

Gauge mediated supersymmetry (SUSY)-breaking (GMSB) furnishes the best idea to overcome the flavor problem in SSMs, supersymmetric stand models (SM). However, implemented in the minimal SSM (MSSM), a very serious fine-tuning problem arises, owing to the absence of a large $A_t$ term in the stop sector to radiatively enhance the SM-like Higgs boson mass. In the extended GMSB coupling $H_u$ to messengers can alleviate this problem but encounters the $A_t/m_{H_u}^2$ problem, i.e., at the same time a large $m_{H_u}^2$ is generated, rendering radiative electroweak spontaneously breaking (EWSB) problematic. This issue shows similarity to another long-standing problem of GMSB, the $\mu/{B_\mu }$ problem, and, of great interest, we find that they may admit the same solution, nonradiative EWSB. Such a solution is naturally accommodated when both $H_u$ and $H_d$ couple to messengers. As a bonus of nonradiative EWSB, the sleptons tend to be very light due to the significant renormalization group equation effect and they are able to account for the $(g-2)_\mu$ puzzle. As a concrete example, we investigate a hidden sector with $(10,\overline {10})$ messengers.

\end{abstract}


\maketitle

\noindent {\bf{Following problems in GSMB and its extension}}  The LHC confirmation of a light spin-0 particle in SM, namely the Higgs boson $h$ having mass $m_h$ around 125 GeV, is a big leap towards the mysterious new physics world, where the gauge hierarchy problem caused by $h$ is supposed to be addressed. Thus far, the most convincing and elegant solution is still SUSY. However, the MSSM, owing to the relative heaviness of $h$ which requires a heavy stop sector to radiatively enhance $m_h$, is suffering a serious little hierarchy problem. Heavy stops push $-m_{H_u}^2\gg \rm(weak~ scale)^2$ via renormalization group equation (RGE) evolutions:
\begin{align}\label{stop:cor}
\Delta m_{H_u}^2\sim -\f{3y_t^2}{4\pi^2}m_{\wt
t}^2\log\f{\Ld_{\rm UV}}{m_{\wt t}}.
\end{align}
$\Ld_{\rm UV}$ is a high scale at which the soft terms are generated and in GMSB it is the messenger scale $M$; $m_{\wt t }$ is the geometric mean of two stop masses, $m_{\wt t_{1}}$ and the heavier one $m_{\wt t_{2}}$. As is well known, Eq.~(\ref{stop:cor}) triggers EWSB radiatively. But now it incurs a serious fine-tuning among parameters in the Higgs sector, manifest in the tadpole equations:
\begin{align}\label{EWSOL1}
\sin2\beta=&\frac{2{B_\mu } }{m_{H_u}^2+m_{H_d}^{2}+2\mu^2},\\
\frac{m_Z^2}{2}=&-\mu^2+\frac{m_{H_d}^2-\tan^2\beta\,
m_{H_u}^2}{\tan^2\beta-1},\label{EWSOL2}
\end{align}%×¢ÒâµÚÒ»¸ö·½³Ì±ÈÕæ¿ÕÓе×ÒªÇó¸üÑϸñ£¨¼´Óм«ÖµÁËÒ»¶¨Óеף©¡£
where $\mu$ is the unique massive coupling of $\mu H_uH_d$ in the MSSM superpotential with unknown origin yet. For $\tan\beta \gg1$ which is good for lifting $m_h$, one can neglect $m_{H_d}^2$ and obtain $-\mu^2 -m_{H_u}^2\approx m_Z^2/2$, so a large cancellation between $\mu^2$ and $m_{H_u}^2$ is required to get the correct weak scale.

In the original GMSB scenario~\cite{early1,early2,early3,GMSB}, that fine-tuning problem is particularly serious~\cite{Draper:2011aa}. Here the stop soft trilinear term $y_tA_t\wt QH_u\wt U^c$ vanishes at the messenger boundary $M$, and then a substantial contribution to $m_h$ from stop mixing, which maximizes for $x_t^2\equiv A^2_t/ m^2_{\wt t }$ near 6~\cite{mms}, is absent. Consequently, one needs $m_{\wt t }\gtrsim 10 \rm TeV$, leading to fine-tuning at the order of magnitude $\sim10^{-4}$ or even worse. Tuning may be substantially alleviated by coupling $H_u$ to messengers through the operator $\ld_uH_u{\cal O}_u$ with ${\cal O}_u$ consisting of messengers only~\cite{Kang:2012ra,Craig:2013wga,Knapen:2013zla}; note that here ${\cal O}_d$ is understood as a set of instead of a single operator. Alternatively, one can couple $Q_3$ and/or $U_3^c$ or even extended matters~\cite{Craig:2012xp,Byakti:2013ti} to messengers.~\footnote{A comparison on the naturalness quality of (some of) the models can be found in Ref.~\cite{Casas:2016xnl}.} In either way, a sizable boundary $A_t$ can be generated and then the need for heavy stops are relaxed.

The former option, the corresponding model dubbed HGMSB, takes the advantage of avoiding recurring of flavor problem. Moreover, it has the potential to be naturally embedded in the solution to another problem of GMSB, namely the  origin of the weak scale $\mu$. Actually, it can be easily generated dynamically by coupling $H_{u}$ and $H_d$ to messengers via $\ld_{u,d}H_{u,d}{\cal O}_{u,d}$, which at one loop gives $\mu\sim\f{\ld_u\ld_d}{ 4\pi } \Ld$ with $\Ld\equiv F/(4\pi M)\sim {\cal O}\rm (10 TeV)$ and $\sqrt{F}$ the hidden sector SUSY-breaking scale. However, the $\mu/B_\mu$ problem follows~\cite{muBmu:early}: Generically $B_\mu$ is also generated at one loop, but it has dimension two and thus is too large to satisfy Eq.~(\ref{EWSOL1}) given that other Higgs parameters have normal sizes. Solving this problem usually involves a quite complicated hidden sector~\cite{muBmu:early,Hall:2002up,Giudice:2007ca}, and in HGMSB one just sets $\ld_d=0$ to avoid it, otherwise complication is indispensable~\cite{Craig:2013wga,Ding:2015vla}.

Even so, it still runs into the EWSB problem. At $M$ a large positive $m_{H_u}^2$ accompanies~\cite{Kang:2012ra} and it cannot be driven negative by RGE (quantities with a hat defined at $M$ otherwise at a scale $\sim m_{\wt t}$),
\begin{align}\label{E:M}
m_{H_u}^2\simeq C_{H_u}\hat m_{H_u}^2-C_{\wt q}\hat m_{\wt q}^2- C_{\wt g}\hat M_3^2>0,
\end{align}
where the coefficients $C_{H_u}$, etc., are positive (see more details around Eq.~(\ref{evolve})); thus, the radiative EWSB fails. This problem is also termed as the $A_t/m_{H_{u}}^2$ problem~\cite{Craig:2013wga}, occurring similarly to the $\mu/B_\mu$ problem: One receives a too large dimension two quantity while generating the desired dimension one quantity, and thus it hampers EWSB. But note that the $A_t/m_{H_{u}}^2$ problem is not as serious as the  $\mu/B_\mu$ problem since unlike $B_\mu$, we will see that the extra $m_{H_u}^2$ term arises at two loop.

\noindent {\bf{Three birds with one stone, nonradiative EWSB}}  Although the HGMSB suffers the above problems, intriguingly they may stand the chance to be addressed simultaneously by carefully reexamining the role of $\ld_dH_d{\cal O}_d$. The key observation is that incorporating it $m_{H_d}^2$, like $\Delta m_{H_u}^2\sim \alpha_u^2 \Ld^2$ (with $\alpha_u\equiv \ld_u^2/4\pi$, a convention for all dimensionless couplings hereafter), also receives a two loop level contribution $\Delta m_{H_d}^2\sim \alpha_d^2 \Ld^2$. Hence, the presence of a very large $B_\mu$ in Eq.~(\ref{EWSOL1}), the situation encountered in solving the $\mu/B_\mu$ problem in the HGMSB, is no longer unacceptable as long as $m_{H_d}^2$ accordingly has a very large size: 
\begin{align}\label{beta}
m_{H_d}^2\simeq 2B_\mu/\sin2\beta\approx \tan\beta B_\mu.
\end{align}
On top of that, the large $m_{H_d}^2$ can overcome a positive $m_{H_u}^2$ to fulfill Eq.~(\ref{EWSOL2}) if $m_{H_d}^2/\tan^2\beta-m_{H_u}^2\sim \mu^2$. Therefore, in principle both the $\mu/B_\mu$ and $A_t/m_{H_u}^2$ problems in HGMSB can be solved coherently, if we give up the traditional radiative EWSB scenario and adopt the nonradiative EWSB scenario, which is triggered by a huge $B_\mu$ but balanced by a huge $m_{H_d}^2$.~\footnote{Nonradiative EWSB for the $\mu/B_\mu$ problem was discussed in Ref.~\cite{Csaki:2008sr} and followed by Ref.~\cite{DeSimone:2011va}, but both used one loop $m_{H_d}^2$, which, however, usually is small; see discussions later.}

Now let us switch to the third bird, addressing the muon anomalous magnetic moment $(g-2)_\mu$ puzzle. There is a longstanding discrepancy between the SM prediction and experimental measurement of $(g-2)_\mu$: $\delta a_\mu=(26.1\pm 8.0)\times10^{-10}$~\cite{g-2}.  To account for it within MSSM, the spectra should contain light (the second family) left-handed sleptons $\wt L_2=(\wt \nu_\mu,\wt\mu_L)$ and not too heavy gauginos; moreover, $\tan\beta$ should be relatively large. For instance, the $\wt\nu_\mu$-wino-Higgsino loop gives~\cite{Cho:2011rk}:
\begin{align}
\delta a_\mu\approx\f{\alpha_2}{2\pi}\f{m_\mu^2 M_{\ld_2}\mu \tan\beta}{m_{\wt \nu_\mu}^4} F_a\L \f{M_{\ld_2}^2}{m_{\wt \nu_\mu}^2},\f{\mu^2}{m_{\wt \nu_\mu}^2}\R,
\end{align}
with $F_a(x,y)$ defined in Ref.~\cite{Cho:2011rk}. 

In our HGMSB, this discrepancy can be marginally filled, as a bonus of implementing nonradiative EWSB. In pure GMSB the quantity ${\cal S}\equiv {\rm Tr}(Y_f \hat m_{\wt f}^2)$ is zero as a result of anomaly cancelation. However, in HGMSB it is nonzero because of the nongauge contributions. In particular, in our framework it has an impressive magnitude and moreover takes a negative sign due to the large positive $\hat m_{H_{d}}^2\sim {\cal O}(10\rm TeV)^2$. As a consequence, during RGE the masses of sparticle with negative hypercharge such as $\wt L$ and $\wt U^c$ may be significantly (and family universally) decreased. So the uncolored $\wt L_2$ with a larger hyper charge tends to be light, making a sizable contribution to $\delta a_\mu$.

As a matter of fact, the RGE reduction on $\wt L$ mass is so significant that it endangers stability of $\wt L$. To avoid tachyonic $\wt L$, one has to impose a strict constraint on the size of $\hat m_{H_{d}}^2$ for a given $M$. In considering this bound, the D-term contribution to slepton masses becomes nonnegligible, and it splits the charged and neutral components of slepton doublets, says $\wt L_2$, by an amount $m_{\wt \nu_\mu}^2-m_{\wt\mu_L}^2=\cos2\beta m_W^2$.

\noindent {\bf{A simple calculable model with $(10, \overline{ 10})$ messengers}} As a concrete example, we follow Ref.~\cite{Kang:2012ra} and consider a model containing messengers $(\Phi,\bar \Phi)$ which form representation $(10,\overline{ 10})$ under $SU(5)-$GUT; the field $10$ (similar for $\overline{10}$) is decomposed into $(Q_\Phi,U_\Phi,E_\Phi)$. Viewing from model building, this representation is economical since it can produce the desired HGMSB structure with minimal messenger types.~\footnote{A substantial cancelation between the Yukawa and gauge contributions lowering down $\Delta m_{H_{u}}^{2}$ (see Eq.~(\ref{Higgs10})) is the main reason for choosing $(10,\overline{ 10})$, but it is unnecessary in nonradiative EWSB and then other representations like $(5,\bar 5)$ reopen.} The model, written in the SM language, takes the form of
 \begin{align}\label{hiddensector}
W_{10}=&\ld_u Q_\Phi H_u U_\Phi+\ld_d\overline Q_\Phi H_d \overline 
U_\Phi+\f{\ld_d'}{2}E_\Phi H_dH_d+\cr &+X\L \ld_Q Q_\Phi\overline Q_\Phi+\ld_U\overline U_\Phi
U_\Phi \R+W_{\rm MSSM},
\end{align}
with $X=M+\theta^2 F$ the SUSY-breaking spurion field. If there are $N_f$ flavor of messengers, the Yukawa coupling matrixes $(\ld_{u,d})_{ij}$ should be understood. In order to reduce parameters, we assume that the Higgs-messenger-messenger couplings respect a $U(N_f)$ flavor symmetry, whereas the third term, crucial in realizing the nonradiative EWSB scenario~\footnote{Although without $\ld_d'$ one still possesses the desired structure of Higgs parameters, we find it does not work quantitatively.}, maximally breaks it; we also assume $(\ld_d')_{i}=\ld_d'$. But general coupling structures do not affect the core of this model. Additionally, gauge invariant allows the $\ld_u'\bar E_\Phi H_uH_u$ term, but we turn it off because of its irrelevance as long as $\ld_u'\ll\ld_u$.

%A larger $N$ will be shown to be helpful in resolving problems. 

Now we present the soft SUSY-breaking parameters from the hidden sector Eq.~(\ref{hiddensector}). The ordinary pure gauge contributions to the gaugino masses $M_{\ld_i} (i=1,2,3)$  and sfermion masses $m_{\wt f}^2$ can be found in Ref.~\cite{GMSB} and are not listed here. Let us focus on the contributions from the Higgs-messenger Yukawa interactions. They also mediate SUSY-breaking effects to the Higgs fields and as well the matter fields which couple to them with sizable strengths, and the resulting soft terms can be extracted utilizing the wave-function renormalization method~\cite{Giudice:1997ni}. For instance, the extra two loop stop/sbottom soft mass squares are
\begin{align}\label{mQ2}
 \Delta m_{\wt Q_3}^{2}=&-\f{N_m}{3}\L 3\alpha_t\alpha_u+3\alpha_b\alpha_d+\alpha_b\alpha_d'\R\Ld^2,\\
 \Delta m_{\wt U_3^c}^2=&-2 N_m \alpha_t\alpha_u\Ld^2,\cr
 \Delta m_{\wt D_3^c}^2=&-2\f{N_m}{3} \alpha_b(3\alpha_d+\alpha_d')\Ld^2.\label{mU2}
\end{align}
Staus get similar contributions but suppressed by $h_\tau^2$. Besides, the desired one loop soft trilinear couplings are 
\begin{align}
A_{t}=-N_m\alpha_u\Ld,~~ A_{b}=A_\tau=-N_m(\alpha_d+\alpha_d'/3)\Ld,
\end{align}
where $N_m\equiv 3N_f$ is the effective messenger index. Requiring that the gauge couplings keep perturbative up to $M_{\rm GUT}$, one gets an upper bound $N_m\lesssim 150/\ln \f{M_{\rm GUT}}{M}$~\cite{GMSB}.

The Higgs-messenger Yukawa interactions have a great impact on the Higgs sector. $H_{u,d}$ couple to messengers directly and their soft mass squares get substantial enhancements for large Higgs-messenger Yukawa couplings:
{\small\begin{align} \label{Higgs10}
\Delta m_{H_{u}}^{2}=N_m{\alpha_u}& \left[(N_m+3)\alpha_u
 \right.
  \nonumber \\ 
&\left.
-\f{16}{3}\alpha_3
-3\alpha_2-\f{13}{15}\alpha_1\right]\Ld^2,\\
\Delta m_{H_{d}}^{2}\approx \f{N_m}{3} \alpha_d' &\L N_m \alpha_d'+2N_m\alpha_d+3\alpha_b-3\alpha_2\R \Ld^2, \label{Higgs10:d} 
\end{align}}There are also one loop contributions. However, they are suppressed by $F^4/M^6\ll1$ due to an accidental cancellation for the single spurion case. But for $\sqrt{F}$ close to $M$, these contributions, always negative, may play an important role in EWSB to solve the $A_t/m_{H_u}^2$ problem~\cite{Craig:2012xp,Jelinski:2015voa}. In our paper this situation should be evaded, otherwise the appealing nonradiative EWSB solution fails. Asides from the soft mass squares, the $\mu/B_\mu$ parameters are also generated at one loop,
\begin{align}\label{mu}
\mu=&f(\ld_Q/\ld_U)N_m \sqrt{\alpha_u\alpha_d} \Ld,\cr
 {B_\mu
}=&f(\ld_Q/\ld_U)N_m4\pi\times \sqrt{\alpha_u\alpha_d}{\Ld^2},
\end{align}
with $f(x)=x \ln x^2/(1-x^2)$. Note that for a hidden sector embedded in GUT, one has $\ld_Q/\ld_U\approx1$ and $f(\ld_Q/\ld_U)$ reaches its maximum 1. Otherwise one may have $f(x)\ll1$, exacerbating the $\mu/B_\mu$ problem.

A large effective messenger number $N_m$ makes good for solving the $\mu/B_\mu$ problem, simply because both $\mu$ and $B_\mu$ are proportional to $N_m$ but they have different dimensions. This is easily seen from the relation $B_\mu=\mu \Ld/4\pi$, where $\Ld$, for a larger $N_m$, now can be much smaller than the single messenger case, thus reducing the gap between $\mu$ and $\sqrt{B_\mu}$. Similarly, stop mixing may benefit from a large $N_m$ since one has $\hat A_t^2/\hat m_{\wt t}^2\propto N$ at the boundary. But in practice we will work in a high messenger scale and thus the significant RGE effects, mainly from gluinos, tend to smear such an effect.

\noindent {\bf{A detailed analysis}} The above UV model defines the low energy MSSM, specifying all of its parameters once the six hidden sector parameters $(\ld_u,\ld_d,\ld_d',N_m,M,\Ld)$ are given. Confronting the $m_h=125$ GeV Higgs boson, it is nontrivial to accommodate all the desired low energy phenomenologies with such a small set of parameters, at least relatively naturally. Before heading towards the detailed analysis of that possibility, we would like to mention that within MSSM the exact calculations of $m_h$ beyond two loop are available and the analytical expression is quite complicated, with a sizable uncertainty~\cite{Draper:2013oza}. But for our purpose it is enough to just output $m_{\wt t}$, $A_t$ and $\tan\beta$, instead of $m_h$ directly; typically, taking into account the theoretical uncertainties, $m_{\wt t}\gtrsim 2$ TeV (with $\tan\beta\gtrsim5$) is still required even near maximal stop mixing but with $A_t<0$, which is just the case in HGMSB.

Now let us enter the detailed analysis. For illustration, we shall employ a semi-analytical method. This is based on the feature that after flowing down to the SUSY scale $M_{\rm SUSY}\equiv M_{S}\simeq1$ TeV, the expressions of the soft terms in general are the combinations of their boundary values~\cite{RGE}:
\begin{align}\label{evolve}
 m_{\wt f}^2\simeq&  C^{\wt f}_{\phi}\hat m_{\phi}^2-C^{\wt f}_{A_t}\hat A_t^2-C^{\wt f}_{ij}\hat M_{\ld_i}\hat M_{\ld_j}-C^{\wt f}_{Ai}\hat A_t \hat M_{\ld_i},\cr
 A_t\simeq&  C_{A_t}\hat A_t-C_{M_i}\hat M_{\ld_i},
\end{align}
where the coefficients $ C^{\wt f}_\phi$ (with $\phi$ denoting the relevant sfermions), asides from $M$, depend on the SM gauge and Yukawa couplings. One can obtain these coefficients numerically~\cite{Kang:2012sy} for a given $M$ like $10^{12}$ GeV, e.g.,
\begin{eqnarray}\label{}
m_{H_u}^2&\approx& 0.63\hat m_{H_u}^2+0.02\hat  m_{H_d}^2-0.36\hat  m_{\wt Q_3}^2-0.31\hat  m_{\wt U^c_3}^2\cr
&-&1.09\hat  M_{\ld_3}^2-0.08\hat  M_{\ld_2}\hat  M_{\ld_3}
+0.20\hat A_t \hat M_{\ld_3}.
\end{eqnarray}

As for the runnings of $\mu$ and $B_\mu$, they are multiplicative for the one loop $\beta$ functions $\beta_{B_\mu}(t)\approx \beta_\mu(t)\approx 3\alpha_t/8\pi $ and drive the boundary values towards the smaller values by a common factor
\begin{align}\label{}
\xi(t)= e^{\int^t_0 \beta_\mu(t') dt' }\lesssim1,
\end{align}
where $t\equiv 2\log \f{Q}{ M}$ with $Q$ the running scale. One gets $\xi_S\equiv \xi(t_S)\approx0. 75(0.85)$ for $M=10^{12}(10^{7})$ GeV and $t_S=\log \f{M_{S}}{M}$. Note that in numerically solving RGEs $\tan\beta$ should be given, whereas it is known just after EWSB and thus iteration in principle is needed for precision. But the Higgs parameters do not depend on $\tan\beta$ to the first approximation for $\tan\beta$ having a normal size, so it is justified to fix $\tan\beta$ like 5. The above treatment is sufficiently good for illustration, and a refined study using expert programs is needed to achieve higher precision.

\begin{figure}
 \begin{center}
 \includegraphics[scale=0.43]{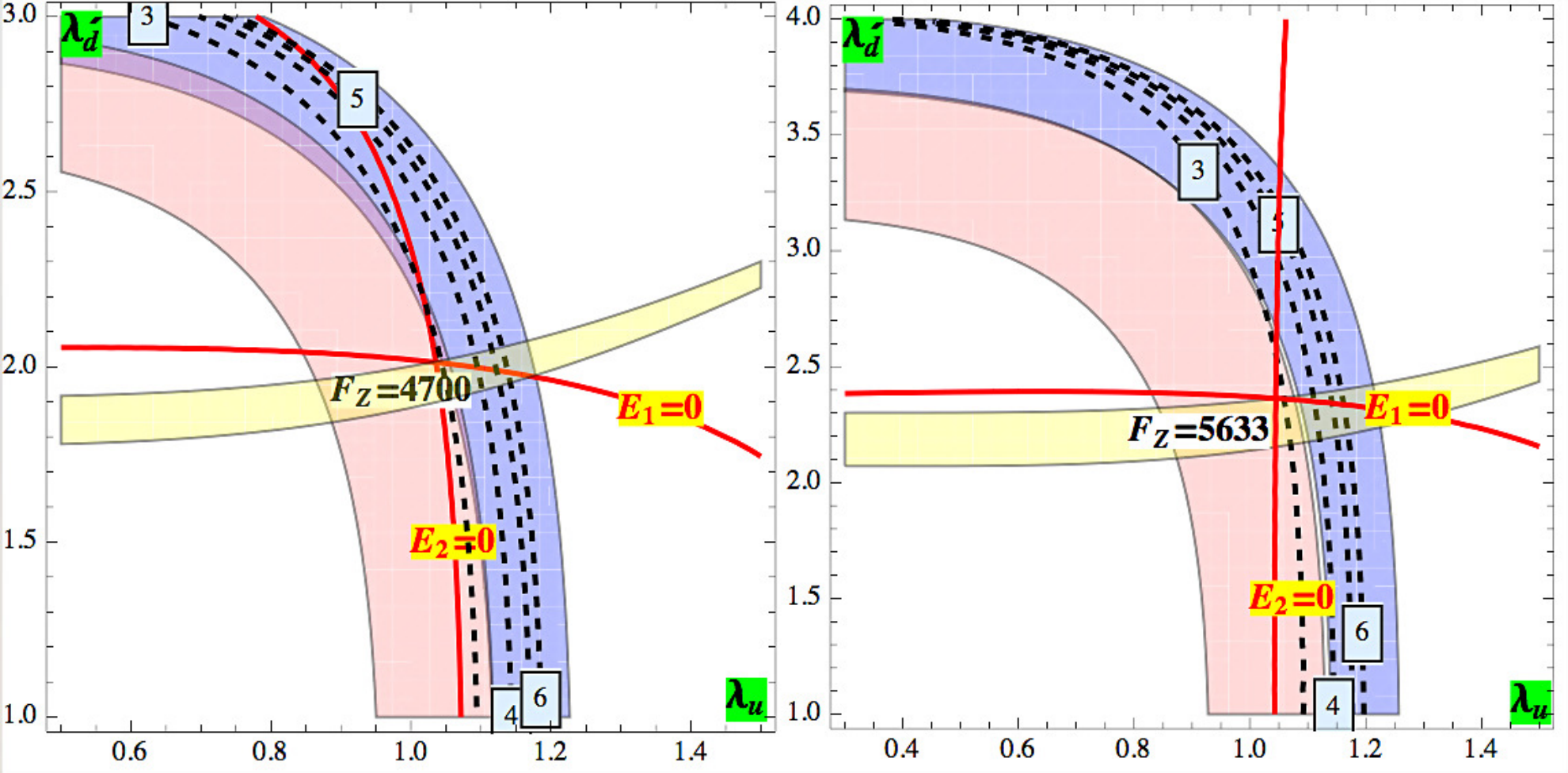}
 \caption{\label{EWSB} Spectra and EWSB on the $\ld_u-\ld_d'$ plane with $M=10^{12} (10^{7})$ GeV on the left (right) panel. The two red lines labelled with $E_1=0$ and $E_2=0$ indicate Eq.~(\ref{EWSOL1}) and Eq.~(\ref{EWSOL2}), respectively, and their intersection is the solution; the dashed lines give value of $x_t^2$. The shaded regions are for $0.2<m_{\wt t_1}/{\rm TeV}<1.5$ (blue), $2<m_{\wt t}/{\rm TeV}<2.5$ (pink) and $0.1<m_{\wt \tau_L}/{\rm TeV}<0.5$ (yellow). To get a smaller fine-tuning $F_Z$, in the left (right) panel we set $N_f=5(2)$, $\tan\beta=13.2(11.1)$, $\mu=110(106)$GeV and $\Ld=2.5(6.0)$TeV. At the two benchmark points, one has $\delta a_\mu=1.4\times10^{-9} (1.3\times 10^{-9})$ and $m_{\wt \nu_\tau}=90.7(82.8)$ GeV, $m_{\wt\tau}=120.9(114.9)$ GeV, $m_{\chi_1}=105.0(100.7)$ GeV and  $m_{\chi_2}=113.0(109.1)$ GeV.  }
 \end{center}
 \end{figure}

With these Higgs parameters at hand, we can investigate EWSB at $M_{S}$. For comparison, we take two typical messenger scales, a relatively high one $M=10^{12}$ GeV and a relatively low one $M=10^{7}$ GeV. $N_m$ will take the maximum allowed by perturbativity. Among the six free parameters, using Eq.~(\ref{beta}), one may fix $\ld_d'$ via
\begin{align}\label{}
\ld_d'\simeq 1.7 \L\f{\xi_S}{0.85}\f{\tan\beta}{5}\f{81}{N_m^2}\R^{\f{1}{4}} \L\f{\mu}{0.2\rm TeV}\f{\rm 10 TeV}{\Ld}\R^{\f{1}{4}},
\end{align}
where we have neglected the small reduction of $m_{H_d}^2$ during RGE; moreover, we just keep the leading term in Eq.~(\ref{Higgs10:d}). This estimation is in good agreement with the numerical solution to tadpole equations shown in Fig.~\ref{EWSB}, on the $\ld_u-\ld'_d$ plane. On this plane, we also demonstrate regions of the lighter stop mass $m_{\wt t_1}$, average stop mass $m_{\wt t}$ and as well contours of stop mixing $x_t^2$, from which we gain knowledge of $m_h$; moreover, a yellow band $100{\rm GeV}<m_{\wt \nu_\mu}<500{\rm GeV}$ is plotted, to indicate the situation of $(g-2)_\mu$, whose values are given in the caption.

Like always, making $m_{\wt t}\gtrsim2$ TeV for Higgs boson mass is the real cause of serious fine-tuning in MSSM. To estimate naturalness of the model, we adopt the conventional Barbieri-Giudice measure~\cite{tuning}:
\begin{align}\label{}
F={\rm Max}_{\ld_i}\left |\f{\partial \log m^2_Z}{\partial \log \ld_i}\right |,
\end{align}
with $\ld_i=(\ld_u,\ld_d,\ld_d',M,N_f,\ld_{\rm SM})$ the set of parameters defined at the boundary $M$. As usual, we start from Eq.~(\ref{EWSOL2}), where $\tan\beta$ has been eliminated after using Eq.~(\ref{beta}). Obviously, the largest fine-tuning, for $\mu$ around the weak scale, arises from the cancelation between the second and third terms in the right-handed side of Eq.~(\ref{EWSOL2}); in other words, here a small $\mu$ does not guarantee naturalness. Fine-tuning is sensitive to the input $\tan\beta$. For a moderate $\tan\beta\sim5$, varying $\ld_u$ leads to the largest tuning $F_Z$ near $10^{4}$. The situation can be improved by increasing $\tan\beta$, which could help to lower down $m_{H_u}^2$ because $m_{H_d}^2/\tan^2\beta\simeq B_\mu/\tan\beta$ decreases. But several reasons hamper a significant improvement in this way. First, a smaller $\ld_u$ incurs tension with maximal stop mixing. Next, increasing $\tan\beta$ actually is at the price of increasing $\ld_d'$ (or $m_{H_d}^2$), and thus it deteriorates the tuning with respect to $\ld_d'$. Last but not least, $\cal S$ becomes so large that  $\wt L$ are driven tachyonic. Therefore, tuning worse than $0.1\%$ is typical, still serious, and the focus point scenario might furnish a way to improve it~\cite{Ding:2013pya}.

%Higgs-messenger couplings and $\Ld$. 

%The sbottom/stau sector also depends on $\tan\beta$, but it happens mainly in the sbottom sector while the 

\noindent {\bf{Light but hidden \& heavy and decoupled}}  Mainly ascribed to the nonradiative EWSB and the attempt for $(g-2)_\mu$ puzzle, the mass spectra shows several remarkable features which yield deep implications to LHC searches for HGMSB of our type. First we outline the heavy spectra. In the Higgs sector, $m_{H_d}\sim 10$ TeV, so the second Higgs doublet whose components have degenerate mass at $m_{H_d}$ is definitely inaccessible at the 14 TeV LHC. In the colored sector, the lighter stop actually is not light, typically near TeV; the $\wt L$ stability bound prevents $\wt t_1$ from being very light. Such a stop is safe under the current LHC exclusions but still promising in the future LHC. Gauginos are also heavy, with bino a few hundred GeVs, since $m_h\simeq125$ GeV requires a high $\Ld$ and moreover we are using a large messenger number $N_m$ which leads to an enhancement of gaugino mass with respect to the sfermion mass by a factor $\sqrt{N_m}$.

Next we move to the very light spectra. $\mu$ should be as small as possible, says just slightly above the LEP II bound around 100 GeV, to achieve a sufficiently large $\delta a_\mu$ (and also a less fine-tuning model). The resulting light Higgsino spectra includes the lightest neutralino $\chi_1$, heavier charginos $\chi^\pm$ and an even heavier neutralino $\chi_2$; their masses are near $\mu$ with small mass splitting 
\begin{align}
\delta m\simeq \f{m_W^2}{M_{\ld_2}^2}(1+\tan^2\theta_wM_{\ld_2}^2/M_{\ld_1}^2)\sim {\cal O}(5)\rm GeV.
\end{align}
For such a compressed Higgsino sector, LHC, given that $\wt\chi_1$ is the next-to-lightest sparticle (NLSP), will not provide sensitivity better than LEP via mono-jet~\cite{Han:2013usa} but can be hunted via some other strategies~\cite{Han:2014kaa,Han:2015lma}. Again to lift $\delta a_\mu$, more likely the NLSP is $\wt \nu_\tau$ (highly degenerate with $\wt \nu_\mu$)~\cite{Katz:2009qx,Guo:2013asa}, having mass just bounded from below by $m_Z/2$. If the slepton spectra is also fairly degenerate with the Higgsino spectra, LHC will be blind to all of them. But if $m_{\wt \chi_2}-m_{\wt \nu_\tau}\sim {\cal O}(50)$ GeV, some Higgsinos may be visible at LHC because the cascade decay of the Higgsinos will produce sufficiently hard multi leptons, e.g., 
\begin{align}
pp\ra \chi^\mp\chi^\pm\ra \tau^+\tau^- +{\rm MET }(=\wt\nu_\tau\wt\nu_\tau^*).
\end{align}
Although a model independent study based on general GMSB has been done in Ref.~\cite{Knapen:2016exe}, an investigation specific to this model is still meaningful; see a relevant study~\cite{Ajaib:2015yma}.

\noindent {\bf{Conclusions and discussions}} HGMSB with $H_{u}$ messenger coupling only can alleviate the little hierarchy problem in GMSB caused by the 125 GeV Higgs boson, but leaving the $A_t/m_{H_u}^2$ problem.  We propose that introducing $H_{d}$ messenger coupling in a proper way can, by means of nonradiative EWSB, address not only this problem but also the $\mu/B_\mu$ problem; moreover, as a bonus the $(g-2)_\mu$ puzzle can be explained. As a concrete example, we consider the hidden sector with $(10,\overline{10})$ messenger but other cases also definitely deserve explorations.

%\noindent {\bf{Acknowledgments:}} Thanks Zhaoxia Heng and Chunli Tong a lot for helpful discussions on the Fortran code Suspect. This work was supported by the National Natural Science Foundation of China under grant Nos. 10821504, 10725526 and 10635030, by the DOE grant DE-FG03-95-Er-40917, and by the Mitchell-Heep Chair in High Energy Physics.

%%%%%%%%%%%%%%%%%%%%

\end{document}